\documentclass[11pt]{article}
\usepackage{epsfig}

\newcommand{\beq}{\begin{equation}}
\newcommand{\eeq}{\end{equation}}
\newcommand{\beqa}{\begin{eqnarray}}
\newcommand{\eeqa}{\end{eqnarray}}

\newcommand{\laem}{\stackrel{<}{\sim}}
\newcommand{\gaem}{\stackrel{>}{\sim}}

\begin{document}

\begin{titlepage}
\def\thepage {}        

\title{Limits on a Composite Higgs Boson}

\author{
R. Sekhar Chivukula\thanks{sekhar@bu.edu}\ \ \& Christian H\"olbling\thanks{hch@bu.edu} \\
{\small\em Department of Physics,
Boston University, Boston, MA 02215, USA.} \\ 
Nick Evans\thanks{n.evans@hep.phys.soton.ac.uk}
\\ {\small\em Department of Physics,
University of Southampton, Southampton, S017 1BJ, UK.} }

\date{March, 2000}

\maketitle

\bigskip
\begin{picture}(0,0)(0,0)
\put(295,250){BUHEP-00-3}
\put(295,235){SHEP-00-02}
\end{picture}
\vspace{24pt}

\begin{abstract}
  Precision electroweak data are generally believed to constrain the
  Higgs boson mass to lie below approximately 190 GeV at 95\%
  confidence level.  The standard Higgs model is, however, trivial and
  can only be an effective field theory valid below some high energy
  scale characteristic of the underlying non-trivial physics.
  Corrections to the custodial isospin violating parameter $T$ arising
  from interactions at this higher energy scale dramatically enlarge
  the allowed range of Higgs mass. We perform a fit to precision
  electroweak data and determine the region in the $(m_H, \Delta T)$
  plane that is consistent with experimental results. Overlaying the
  estimated size of corrections to $T$ arising from the underlying
  dynamics, we find that a Higgs mass up to 500 GeV is allowed.  We
  review two composite Higgs models which can realize the possibility
  of a phenomenologically acceptable heavy Higgs boson.  We comment on
  the potential of improvements in the measurements of $m_t$ and $M_W$ to
  improve constraints on composite Higgs models.
\pagestyle{empty}
\end{abstract}
\end{titlepage}

\setcounter{section}{1}

\setcounter{equation}{0}

Precision electroweak data are generally believed to constrain the
Higgs boson mass to lie below approximately 190 GeV at 95\% confidence
level \cite{Moriondewwg,LEPEWWG}.  The standard Higgs model is,
however, trivial \cite{Wilson} and
can only be an effective field theory valid below some high energy
scale $\Lambda$ characteristic of the underlying non-trivial physics.
Additional interactions coming from the underlying theory, and
suppressed by the scale $\Lambda$, give rise to model-dependent
corrections to measured electroweak quantities. When potential
corrections from physics at higher energy scales are included, the
limit on the Higgs boson mass becomes weaker\footnote{See also,
  Langacker and Erler in \protect\cite{Caso:1998tx}.}
\cite{Alam,Chivukula:1999az}.

In the context of the triviality of the standard model and given the
relatively weak (logarithmic) dependence of electroweak observables on
the Higgs boson mass \cite{Hagiwara:1997zm}, the
typical size of corrections to $T$ arising from custodial symmetry
violating \cite{Weinstein:1973gj} non-trivial
underlying physics can dramatically enlarge\footnote{In contrast, in
  theories lacking a custodial symmetry the contributions to $S$ are
  relatively small
  \protect\cite{Chivukula:1999az,Chivukula:1996sn}
  and do not have a significant effect on Higgs mass bounds.}  the
allowed Higgs mass range \cite{Chivukula:1999az}. In this note we
perform a fit to precision electroweak data and determine the region
in the $(m_H, \Delta T)$ plane that is consistent with experimental
results. Overlaying the predicted size of corrections to $T$ arising
from the underlying dynamics, we find that a Higgs mass up to 500 GeV
is allowed. We review two composite Higgs models which can realize the
possibility of a phenomenologically acceptable heavy Higgs boson.

For a given Higgs boson mass, an upper bound on the scale $\Lambda$ is
given by the position of the Landau pole \cite{Dashen:1983ts} of the
Higgs boson self-coupling $\lambda$. As the Higgs boson mass is
proportional to $\lambda(m_H)$, the larger the Higgs boson mass the
smaller the upper bound on the scale $\Lambda$. We may
estimate\footnote{While this estimate is based on perturbation theory,
  non-perturbative calculations yield essentially the same result
  \protect\cite{Kuti:1988nr}.}
this upper bound by integrating the one-loop beta function for the
self-coupling $\lambda$, which yields \cite{Dashen:1983ts}
\beq
\Lambda \laem m_H \exp\left({4\pi^2v^2\over 3m^2_H}\right)~,
\label{landau}
\eeq
where $m_H$ is the Higgs boson mass and $v \approx 246$ GeV is the
vacuum expectation value of the Higgs boson.

The leading corrections to electroweak observables from the underlying
theory are encoded in dimension six operators
\cite{Appelquist:1980ix}
which contribute to the Peskin-Takeuchi $S$ and $T$ parameters
\cite{Peskin:1990zt}. Given the scale of the underlying
non-trivial physics, dimensional analysis
\cite{Weinberg:1979kz} may be used to
estimate the size of effects from these dynamics in the low-energy
Higgs theory \cite{Chivukula:1996sn}. If the
underlying theory does not respect custodial symmetry
\cite{Weinstein:1973gj}, the contribution
to $T$ is dominant and is estimated to be
\beq
|\Delta T| \simeq {b\kappa^2\, v^2 \over \alpha_{em}(M^2_Z)\, \Lambda^2},
\label{testimate}
\eeq
or larger \cite{Weinberg:1979kz}.  Here $\alpha_{em}$
is the electromagnetic coupling renormalized at $M^2_Z$, $b$ is a
model-dependent coefficient of order 1, and $\kappa$ is a measure of
the size of dimensionless couplings in the effective Higgs theory
and is expected to lie between 1 and $4\pi$ \cite{Weinberg:1979kz}.
Combining eqn. \ref{testimate} with the bound on $\Lambda$ shown
in eqn. \ref{landau}, we find
\beq
|\Delta T| \gaem {b\kappa^2\, v^2 \over \alpha_{em}(M^2_Z)\, m^2_H}\,
\exp\left({-\,{8 \pi^2 v^2\over 3 m^2_H}}\right)~.
\label{tmhestimate}
\eeq

Since the Higgs model is trivial, the potential effects of the
underlying non-trivial dynamics must be included when establishing
constraints on the Higgs mass \cite{Chivukula:1999az}. As the
contributions to $T$ are expected to dominate, we have performed a fit
to electroweak measurements \cite{LEPEWWG} and have determined the
region in the $(m_H, \Delta T)$ plane that is consistent with these
results. In addition to measurements at the $Z$-pole from LEP and SLD,
we include measurements of $M_W$ from LEP and the Tevatron,
and measurements of $m_t$ from the Tevatron.  In performing these
fits, we have used ZFITTER 6.21 \cite{Bardin:1999yd} to generate the
standard model predictions for a given value of the $Z$ mass, Higgs
mass, top-quark mass, and strong ($\alpha_s$) and electromagnetic
($\alpha_{em}$) couplings, and have introduced the effect of non-zero
$\Delta T$ linearly \cite{Burgess:1994vc}.  We have included the
determinations of $\alpha_{em}$
\cite{Alemany:1997tn}
\beq
\alpha^{-1}_{em}(M^2_Z) = 128.905 \pm 0.036
\eeq
and the (non-electroweak) determinations of $\alpha_s$
\cite{Caso:1998tx}
\beq
\alpha_s(M^2_Z) = 0.119 \pm 0.002
\label{alphas}
\eeq
as observations, {\it i.e.} we have included deviations from the
listed central values in our computation of $\chi^2$. The correlation
matrices listed in ref. \cite{LEPEWWG} are incorporated in our
calculation of $\chi^2$.

\begin{figure}[tbp]
\begin{center}
\includegraphics*[width=12cm]{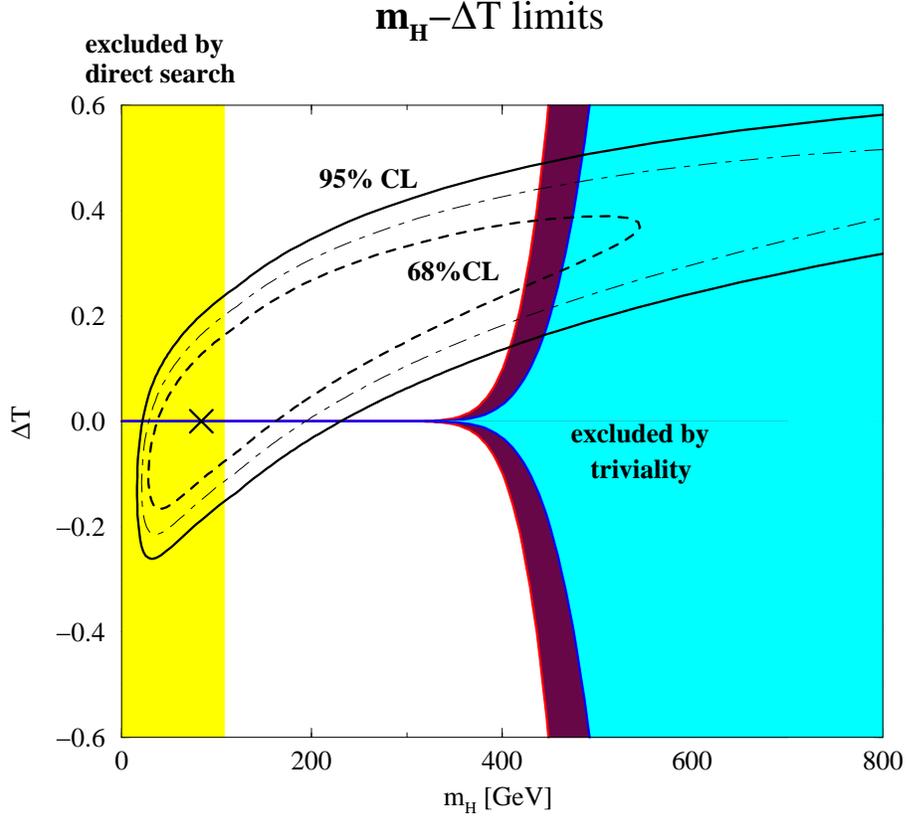}
\caption
{68\% and 95\% CL bounds in the $(m_H,\Delta T)$ plane allowed by a
  fit to precision electroweak data
  \protect\cite{Moriondewwg,LEPEWWG}. The best fit ``standard model''
  value is shown by the cross on the $\Delta T=0$ line. (Also shown by
  the dot-dash curve is the contour corresponding to $\Delta \chi^2=4$,
  whose intersection with the line $\Delta T=0$ -- at approximately 190
  GeV -- corresponds to the usual 95\% CL upper bound quoted on the
  Higgs boson mass in the standard model.) The light region to the
  right is excluded by eqn.  \ref{tmhestimate} for $b\kappa^2 = 4\pi$.
  The dark region denotes the additional area excluded for
  $b\kappa^2=4\pi^2$. The positive branches of the curves bounding
  these regions are lower bounds for $\Delta T$ in the top-seesaw and
  composite higgs models described in the text. Any $(m_H, \Delta T)$
  with positive $\Delta T$ and to the left of the appropriate
  triviality curve can be realized in the corresponding model.}
\label{Fig1}
\end{center}
\end{figure}

The result of our fit is summarized in Figure \ref{Fig1}. The best-fit
value\footnote{This best fit value is, of course, below the direct
  experimental lower bound \protect\cite{Moriond00} of order 108 GeV.}
is shown and occurs at a Higgs boson mass of 90 GeV; it corresponds to
a minimum value of $\chi^2 = 21.7$ for 21 observables while varying 5
fit parameters ($M_Z$, $m_H$, $m_t$, $\alpha_s$, and $\alpha_{em}$).
For two degrees of freedom, the 68\% and 95\% CL bounds correspond to
$\Delta \chi^2$ of 2.30 and 6.17 respectively. The two
degree\footnote{Note that the one degree of freedom 95\% CL upper
  bound on $m_H$, $\Delta \chi^2=4$ and $\Delta T=0$, is approximately 190 GeV
  in agreement with \protect\cite{Moriondewwg}.} of freedom 95\% CL
upper bound on $m_H$ is 243 GeV for $\Delta T=0$.

Extending this bound to non-zero $\Delta T$, we see that the region in
the $(m_H, \Delta T)$ plane which fits the observed data as well as
the ``standard model'' at 95\% CL extends to large Higgs masses for a
positive value of $\Delta T$.

It is not possible, however, for this {\it entire} region to be
realized consistent with the constraints of triviality. For example,
motivated by the models we consider below, the area excluded by eqn.
\ref{tmhestimate} with $b\kappa^2 = 4\pi$ is shown as the light region
on the right in Figure \ref{Fig1}.  Overlaying the constraints, we see
that Higgs masses above 500 GeV would likely imply the existence of
new physics at such low scales ($\Lambda \laem 12$ TeV from eqn.
\ref{landau}) as to give rise to a contribution to $T$ which is too
large \cite{Chivukula:1996sn}.  

We emphasize that these estimates are based on dimensional arguments,
and we are not arguing that it is impossible to construct a composite
Higgs model consistent with precision electroweak tests with $m_H$
greater than 500 GeV.  Rather, barring accidental cancellations in a
theory without a custodial symmetry, contributions to $\Delta T$
consistent with eqn. \ref{tmhestimate} are generally to be expected.
This expectation is illustrated in the two models which we now
discuss.

The top-quark seesaw theory of electroweak symmetry breaking
\cite{Dobrescu:1998nm,Collins:1999rz} provides a
simple example of a model with a potentially heavy composite Higgs
boson consistent with electroweak data. In this case, electroweak
symmetry breaking is due to the condensation, driven by a strong
topcolor \cite{Hill:1991at} gauge interaction, of the left-handed
top-quark with a new right-handed singlet fermion $\chi$. Such an
interaction gives rise to a composite Higgs field at low energies, and
the mass of the top-color gauge boson sets the scale of the Landau
pole $\Lambda$ \cite{Bardeen:1990ds}.  The weak singlet $\chi_L$ and
$t_R$ fields are introduced so that the $2\times 2$ mass matrix,
\beq
\left(
\begin{array}{cc}
0 & m_{t\chi} \\
m_{\chi t} & m_{\chi \chi}
\end{array}
\right)
\eeq
is of seesaw form and has a light eigenvalue corresponding to the
observed top quark. The value of $m_{t \chi}$ is related to the weak
scale, and its value is estimated to be 600 GeV
\cite{Dobrescu:1998nm}.

The coupling of the top-quark to $\chi$ violates custodial symmetry
in the same way that the top-quark mass does in the standard model.
The leading contribution to $T$ from the underlying top seesaw physics
arises from contributions to $W$ and $Z$ vacuum polarization diagrams
involving the $\chi$. This contribution is positive and is calculated to be
\cite{Dobrescu:1998nm,Collins:1999rz}
\beq
\Delta T = {N_c \over 16 \pi^2 \alpha_{em}(M^2_Z) }\, {m^4_{t\chi}\over m^2_{\chi\chi} v^2}
\approx {0.7\over \alpha_{em}} \left({\Lambda^2 \over m^2_{\chi\chi}}\right)\,
\left({v^2\over \Lambda^2}\right)~,
\eeq
which is of the form of eqn. \ref{testimate} with $b\kappa^2 \propto
(\Lambda/m_{\chi\chi})^2$.  Note that $\Lambda/m_{\chi\chi}$ {\it
  cannot} be small since top-color gauge interactions must drive
$t\chi$ chiral symmetry breaking. Taking $\Lambda/m_{\chi\chi}\approx
4$, we reproduce the positive branch of the boundary of the light
region excluded by triviality shown in Figure \ref{Fig1}. By varying
$\Lambda$ and $m_{\chi\chi}$, the entire allowed $(m_H,\Delta T)$
region with positive $\Delta T$ and to the left of the triviality
constraint can be obtained.  In particular, we note that it is
possible to obtain a {\it light} Higgs boson in this context as well.

The fact that contributions to $T$ greatly expand the region of
allowed Higgs mass in the top seesaw model is discussed in detail in
ref.  \cite{Collins:1999rz}. Here we see that the running of the Higgs
self-coupling encoded in the constraints of eqn.  \ref{tmhestimate}
prevent Higgs masses higher than about 500 GeV from being realized
\cite{Chivukula:1996sn}.

``Composite Higgs Models''
\cite{Kaplan:1984fs}
also provide examples of theories with a potentially heavy composite
Higgs boson.  In the simplest of these models, one introduces three
new fermions which couple to a vectorial ``ultracolor'' $SU(N)$ gauge
interaction.  Two of these fermions ($\psi$) transform as a vectorial
doublet under $SU(2)_W$, while the third ($\sigma$) is assumed to be a
singlet.  Dirac mass terms can be introduced for all of these fermions
and, as so far described, chiral symmetry breaking driven by the
ultracolor interactions leaves the vectorial $SU(2)_W \times U(1)_Y$
unbroken. Extra chiral interactions are then introduced to misalign
the vacuum by a small amount, causing a nonzero $\bar{\psi} \sigma$
condensate and breaking the weak interactions. 

The octet of pions which result from ultracolor chiral symmetry
breaking include a set, the analogs of the kaons, which form a
composite Higgs boson. Models can be constructed \cite{Maekawa:1995yd}
in which the Higgs boson can formally be as heavy as a TeV ({\it i.e.}
at tree-level), while the other four pions have masses controlled by
the ultracolor scale and can be much heavier.

This simplest model does not have a custodial symmetry.
A direct calculation of the $W$ and $Z$ masses yields 
the positive contribution
\beq
\Delta T = {v^2 \over 4\, \alpha_{em}(M^2_Z)\, f^2}~.
\label{compositehiggs}
\eeq
Here $f$ is the pion decay constant for ultracolor chiral symmetry breaking,
the analog of $f_\pi$ in QCD. The ultracolor chiral symmetry breaking scale,
estimated \cite{Weinberg:1979kz} to be ${\cal
  O}(4\pi f)$, sets the compositeness scale $\Lambda$ of the Higgs boson.
Comparing eqns. \ref{testimate} and \ref{compositehiggs}, we see that the
contribution to $T$ is of the same form with $b \kappa^2 \approx 4 \pi^2$,
excluding the light and dark shaded regions to the right in Figure
\ref{Fig1}. From this we see that phenomenologically acceptable composite
Higgs models can be constructed with Higgs masses up to approximately 450
GeV. Again, in this case by varying the Dirac masses of the fermions and
adjusting the size of the chiral interaction, it is possible to construct
models that realize any $(m_H,\Delta T)$ to the left of the triviality
constraint for positive $\Delta T$.

\begin{figure}[tbp]
\begin{center}
\includegraphics*[width=12cm]{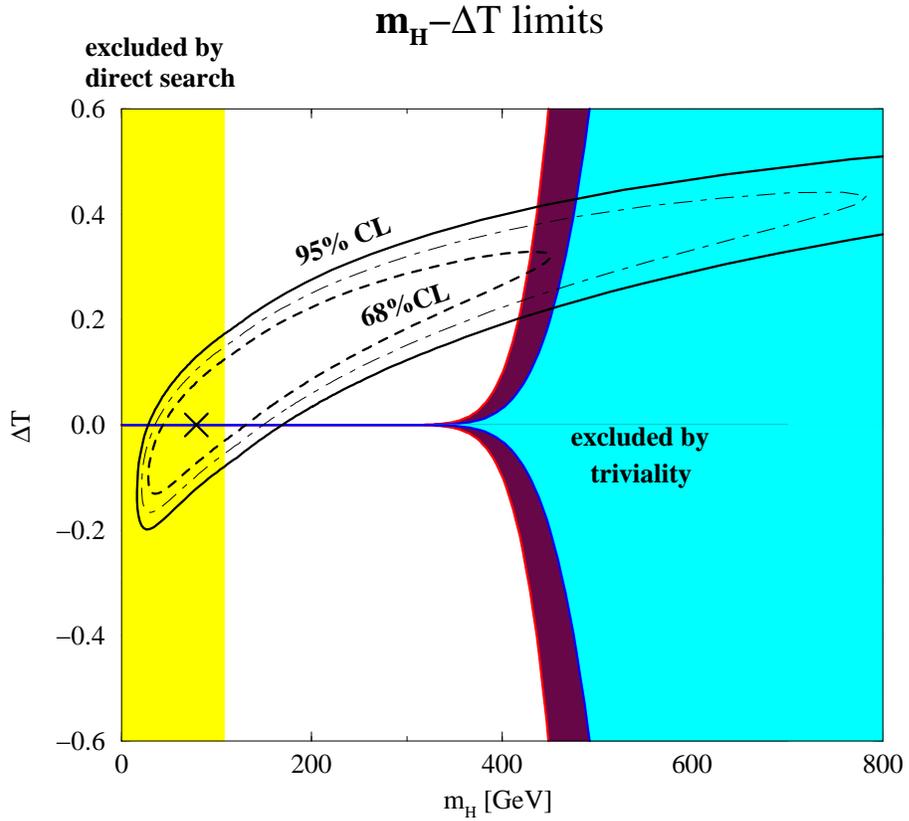}
\caption
{Prospective 68\% and 95\% CL bounds in $(m_H,\Delta T)$ plane allowed
  by fit to precision electroweak data \protect\cite{LEPEWWG} assuming
  uncertainty in $M_W$ is reduced to 30 MeV and uncertainty in $m_t$
  is reduced 2 GeV, as may be possible during Run II of the Fermilab
  Tevatron. All curves are as described in Figure 1.}
\label{Fig2}
\end{center}
\end{figure}

Finally, we briefly consider the prospects for improving these
indirect limits over the next few years. The measurements of $M_W$ and
$m_t$ are likely to be greatly improved during Run II of the Fermilab
Tevatron.  With an integrated luminosity of 10 fb$^{-1}$, it may be
possible to reduce the uncertainty in the top mass to 2 GeV and in the
$W$ mass to 30 MeV \cite{TEV2000}. To illustrate the potential of
these measurements, in Figure \ref{Fig2} we plot the 68\% and 95\% CL
bounds in the $(m_H,\Delta T)$ plane which would be allowed if $M_W$
and $m_t$ assumed their current ``best-fit'' values while the
uncertainties dropped as projected. Note that although the 95\% CL
region is somewhat smaller than in Fig. \ref{Fig1} ({\it e.g.} the two
degree of freedom upper bound on the ``standard model'' Higgs boson
mass -- $\Delta T=0$ -- drops to ${\cal O}$(180 GeV)), there would
still be composite Higgs models consistent with electroweak data with
a Higgs boson mass up to 500 GeV for positive $\Delta T$.

In a forthcoming publication \cite{unpublished}, we will detail the
calculation of corrections to precisely measured electroweak
quantities in the two composite Higgs models we reviewed above and
consider the complementary constraints arising from bounds on $Z \to
b\bar{b}$, flavor-changing neutral currents \cite{Chivukula:1996sn},
and CP-violation.


\centerline{\bf Acknowledgments}

We thank Gustavo Burdman, Aaron Grant, Marko Popovic, and Elizabeth
Simmons for useful discussions.  NE is grateful to PPARC for the
sponsorship of an Advanced Fellowship.  {\em This work was supported
  in part by the Department of Energy under grant DE-FG02-91ER40676.}



\end{document}